\documentclass[prd]{revtex4}
\usepackage{amsmath}
\usepackage{graphicx}
\usepackage{amsfonts}
\usepackage{amssymb}

\begin{document}
\title{Resonant production of diquarks at high energy $pp$, $ep$ and $e^{+}e^{-}$ colliders}
\author{O. \c{C}ak{\i}r}
\email{ocakir@science.ankara.edu.tr}
\author{M. \c{S}ahin}
\affiliation{Ankara University, Faculty of Sciences, Department of Physics, 06100,
Tandogan, Ankara, Turkey}

\begin{abstract}
Resonant productions of the first generation scalar and vector diquarks at
high energy hadron-hadron ($pp$), lepton-hadron ($ep$) and lepton-lepton
($e^{+}e^{-}$) colliders are investigated. Taking into account the hadronic
component of the photon, diquarks can be produced resonantly in the
lepton-hadron and lepton-lepton collisions. Production rates, decay widths and
signatures of diquarks are discussed using the general, $SU(3)_{C}\times
SU(2)_{W}\times U(1)_{Y}$ invariant, effective Lagrangian. The corresponding
dijet backgrounds are examined in the interested invariant mass regions. The
attainable mass limits and couplings are obtained for the diquarks that can be
produced in hadron collisions and in resolved photon processes. It is shown
that hadron collider with center of mass energy $\sqrt{s}=14$ TeV will be able
to discover scalar and vector diquarks with masses up to $m_{DQ}$=9 TeV for
quark-diquark-quark coupling $\alpha_{DQ}$=0.1. Relatively, lighter diquarks
can be probed at $ep$ and $e^{+}e^{-}$ collisions in more clear environment.

\end{abstract}
\maketitle

\section{Introduction}

Diquarks can occur in many scenarios which involve new physics beyond the
standard model (SM), e.g., composite models \cite{Wudka86} and
superstring-inspired $E_{6}$ models \cite{Hewett89}. These particles can
transform as anti-triplet ($3^{\ast}$) or sextet ($6$) under $SU(3)$. Diquarks
carry baryon number $|B|=2/3$ and couple to a pair of quarks. They have
integer spin (scalar or vector) and have electric charges $|Q|=1/3,$ $2/3$ or
$4/3$.

The Collider Detector at Fermilab (CDF) has set limits on the masses of a
class of scalar diquarks decaying to dijets with the exclusion of mass range
$290<m_{DQ}<420$ GeV \cite{Dorigo97}, which are expected to be approximately
valid for other scalar diquarks. There are also indirect bounds imposed on
couplings from electroweak precision data \cite{Bhattacharyya95} from LEP
$e^{+}e^{-}$ collider where these bounds allow diquark-quark couplings up to a
value $\alpha_{DQ}\sim0.1$.

\begin{table}[ptb]
\caption{The main parameters of the future $ep$ and $e^{+}e^{-}$ colliders,
L$^{int}$ denotes the integrated luminosity for one working year.}%
\label{table1}
\begin{center}
$%
\begin{array}
[c]{ccccc}\hline
ep\text{ Colliders} & E_{e}\text{(TeV)} & E_{p}\text{(TeV)} & \sqrt{s_{ep}%
}\text{(TeV)} & L_{ep}^{\text{int}}\text{(10}^{2}\text{pb}^{-1}\text{)}%
\\\hline
\text{ILC}\times\text{LHC} & 0.25 & 7 & 2.64 & 1-10\\
\text{CLIC}\times\text{LHC} & 0.5 & 7 & 3.74 & 1-10\\
\text{CLIC}\times\text{LHC} & 1.5 & 7 & 6.48 & 1-10\\\hline
e^{-}e^{+}\text{ Colliders} & E_{e^{+}}\text{(TeV)} & E_{e^{-}}\text{(TeV)} &
\sqrt{s_{e^{+}e^{-}}}\text{(TeV)} & L_{e^{+}e^{-}}^{\text{int}}\text{(10}%
^{5}\text{pb}^{-1}\text{)}\\\hline
\text{ILC} & 0.25 & 0.25 & 0.5 & 1-10\\
\text{CLIC} & 0.5 & 0.5 & 1.0 & 1-10\\
\text{CLIC} & 1.5 & 1.5 & 3.0 & 1-10\\\hline
\end{array}
$
\end{center}
\end{table}

Three types of the colliders related to the energy frontiers in particle
physics research seem to be promising in the next decade. Namely, they are
Large Hadron Collider (LHC) with the center of mass energy $\sqrt{s}$=14 TeV
and luminosity L$=10^{34}-10^{35}$ cm$^{-2}$s$^{-1}$, International Linear
Collider (ILC) with $\sqrt{s}=0.5$ TeV and L$=10^{34}-10^{35}$ cm$^{-2}%
$s$^{-1},$ Compact Linear Collider (CLIC) with $\sqrt{s}=3$ TeV and
L$=10^{34}-10^{35}$ cm$^{-2}$s$^{-1}$ in the most preferable design, and the
linac-ring type $ep$ colliders, when the linear collider is constructed near
the proton ring, i.e., ILC$\otimes$LHC based $ep$ collider with $\sqrt
{s}=2.64$ TeV and CLIC$\otimes$LHC based $ep$ collider with $\sqrt{s}=3.74$
TeV or $\sqrt{s}=6.48$ TeV, having a luminosity L$=10^{31}-10^{32}$ cm$^{-2}%
$s$^{-1}$. Even though the last one has a lower luminosity it can provide
better conditions for investigations of a lot of phenomena comparing to ILC
due to the essentially higher center of mass energy and LHC due to more clear
environment. The high energy linear $e^{+}e^{-}$ colliders have been proposed
as the instruments that can perform precision measurements that would
complement those performed at the LHC. The diquarks are expected to be easily
observable at LHC through their resonant production and subsequent decay to
two jets. If relatively light diquarks are observed at LHC they are expected
to be in the reach of the future linear $e^{+}e^{-}$ and linac-ring type $ep$
colliders. The main parameters of future $ep$ and $e^{+}e^{-}$ colliders are
given in Table \ref{table1}.

The production and possibility of detection of diquarks have been analysed for
$e^{+}e^{-}$ \cite{Gusso04}, $p\overline{p}$ \cite{Angelopoulos87,Argyres88}
and $pp$ \cite{Atag98,Arik02} colliders. The single and pair production of
scalar diquarks in $ep$ collisions have been analysed in \cite{Rizzo89}
without taking into account hadronic structure of the photon. Although the
photon is the gauge particle of the electromagnetic interactions and thus
pointlike, it is known to behave like a hadron if it interacts with other
hadrons. This can be described by the QCD-corrected quark parton model if the
photon is probed at a large momentum scale. The importance of resolved photon
contributions has been demonstrated by the DELPHI \cite{DELPHI} and OPAL
\cite{OPAL} Collaborations in obtaining interesting limits on leptoquark
properties from $e\gamma$ production of leptoquarks \cite{Doncheski}. The
relevance of the photon substructure increases with increasing center of mass
energy and becomes therefore important for forthcoming $ep$ and $e^{+}e^{-}$ colliders.

\begin{figure}[ptb]
\begin{center}
\includegraphics[width=15cm,height=4cm] {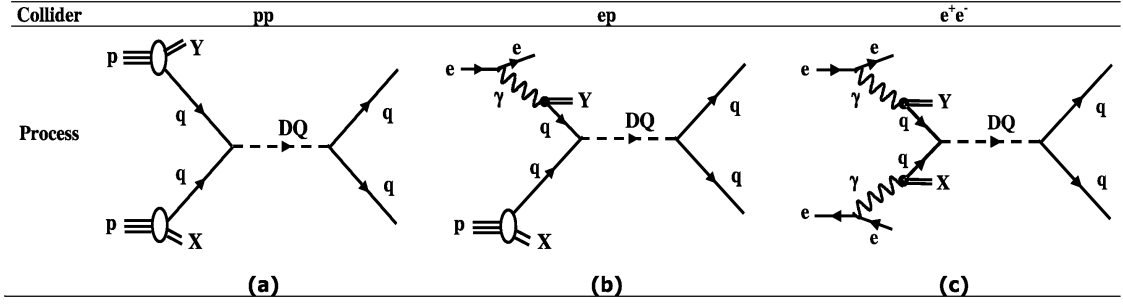}
\end{center}
\caption{Schematic presentation of resonant production of scalar or vector
diquarks at three types of colliders: a) $pp$ collisions, b) $ep$ collisions
with one resolved photon process and c) $e^{+}e^{-}$ collisions with two
resolved photon process. }%
\label{fig1}%
\end{figure}

In this work, we investigate the potentials of high energy $pp$, $ep$ and
$e^{+}e^{-}$ colliders to search for scalar and vector diquarks via their
resonant production subprocesses%

\begin{align}
qq  &  \rightarrow DQ\rightarrow2j\\
q_{\gamma}q  &  \rightarrow DQ\rightarrow2j\\
q_{\gamma}q_{\gamma}  &  \rightarrow DQ\rightarrow2j
\end{align}
where $q_{\gamma}$ is the quark from resolved photon. The second process can
also be considered as a contribution to the single diquark production at ep
colliders. We analyze the relevant background processes in the cases of
three-type of collisions. Schematic presentation of resonant production of
diquarks at three types of colliders is shown in Fig. \ref{fig1} (a-c).

\section{Interaction Lagrangian}

A model independent, baryon number conserving, most general $SU(3)_{C}\times
SU(2)_{W}\times U(1)_{Y}$ invariant effective lagrangian for scalar and vector
diquarks has the form \cite{Atag98,Arik02}
\begin{align}
L_{|B|=0}  &  =f_{1L}\overline{q}_{L}\gamma^{\mu}q_{L}DQ_{1\mu}^{c}%
+(f_{1R}\overline{d}_{R}\gamma^{\mu}d_{R}+f_{1R}^{\prime}\overline{u}%
_{R}\gamma^{\mu}u_{R})DQ_{1\mu}^{\prime c}\nonumber\\
&  +\widetilde{f}_{1R}\overline{u}_{R}\gamma^{\mu}d_{R}\widetilde{DQ}_{1\mu
}^{c}+f_{3L}\overline{q}_{L}\mathbf{\tau}\gamma^{\mu}q_{L}\cdot\mathbf{DQ}%
_{3\mu}^{c}\nonumber\\
&  +f_{2}\overline{q}_{L}i\tau_{2}u_{R}DQ_{2}^{c}+\widetilde{f}_{2}%
\overline{q}_{L}i\tau_{2}d_{R}\widetilde{DQ}_{2}^{c}+\text{H.c.}\label{4}\\
& \nonumber\\
L_{|B|=2/3}  &  =(g_{1L}\overline{q}_{L}^{c}i\tau_{2}q_{L}+g_{1R}\overline
{u}_{R}^{c}d_{R})DQ_{1}^{c}+\widetilde{g}_{1R}\overline{d}_{R}^{c}%
d_{R}\widetilde{DQ}_{1}^{c}\nonumber\\
&  +\widetilde{g}_{1R}^{\prime}\overline{u}_{R}^{c}u_{R}\widetilde{DQ}%
_{1}^{\prime c}+g_{3L}\overline{q}_{L}^{c}i\tau_{2}\mathbf{\tau}q_{L}%
\cdot\mathbf{DQ}_{3}^{c}\nonumber\\
&  +g_{2}\overline{q}_{L}^{c}\gamma^{\mu}d_{R}DQ_{2\mu}^{c}+\widetilde{g}%
_{2}\overline{q}_{L}^{c}\gamma^{\mu}u_{R}\widetilde{DQ}_{2\mu}^{c}+\text{
H.c.} \label{5}%
\end{align}
where the $|B|=0$ diquarks are familiar fields, as they resemble the
electroweak gauge vectors, and the neutral and charged Higgs scalars. Here, we
consider only the $|B|=2/3$ diquarks. In Eq. (4) and (5), $q_{L}=(u_{L}%
,d_{L})$ denotes the left handed quark spinor and $q^{c}=C\bar{q}^{T}$
($\bar{q}^{c}=-q^{T}C^{-1}$) is the charge conjugated quark field. Following
\cite{Atag98,Arik02}, the possible scalar and vector diquarks are $SU(3)_{C}$
anti-triplets and sextets. For the sake of simplicity, color and generation
indices are ommitted in (4) and (5). Scalar diquarks $DQ_{1}$, $\widetilde
{DQ}_{1}$, $\widetilde{DQ}_{1}^{\prime}$ are $SU(2)_{W}$ singlets and
$\mathbf{DQ}_{3}$ is $SU(2)_{W}$ triplet. Vector diquarks $DQ_{2}$ and
$\widetilde{DQ}_{2}$ are $SU(2)_{W}$ doublets. At this stage, we assume that
each SM generation has its own diquarks and couplings in order to avoid
flavour changing neutral currents (FCNC). Furthermore, the members of a given
multiplet are assumed to be the mass degenerated. A general classification of
the first generation, color anti-triplet ($\mathbf{3}^{\ast}\mathbf{)}$
diquarks is shown in Table \ref{table2}.

\begin{table}[ptb]
\caption{Quantum numbers of the first generation, color $\bar{3}$ diquarks
described by the effective lagrangian in the text according to $SU(3)_{C}%
\times SU(2)_{W}\times U(1)_{Y}$ invariance where the hypercharge
$Y=2(Q-I_{3})$.}%
\label{table2}
\begin{center}%
\begin{tabular}
[c]{cccccc}
&  &  &  &  & \\\hline
& SU(3)$_{C}$ & SU(2)$_{W}$ & U(1)$_{Y}$ & $Q$ & Couplings\\\hline
Scalar Diquarks &  &  &  &  & \\
$DQ_{1}$ & 3$^{\star}$ & 1 & 2/3 & 1/3 & $u_{L}d_{L}(g_{1L}),\,u_{R}%
d_{R}(g_{1R})$\\
$\widetilde{DQ}_{1}$ & 3$^{\star}$ & 1 & -4/3 & 2/3 & $d_{R}d_{R}(\tilde
{g}_{1R})$\\
$\widetilde{DQ}_{1}^{\prime}$ & 3$^{\star}$ & 1 & 8/3 & 4/3 & $u_{R}%
u_{R}(\tilde{g}_{1R}^{\prime})$\\
$DQ_{3}$ & 3$^{\star}$ & 3 & 2/3 & $\left(
\begin{array}
[c]{c}%
4/3\\
1/3\\
-2/3
\end{array}
\right)  $ & $\left(
\begin{array}
[c]{c}%
u_{L}u_{L}(\sqrt{2}g_{3L})\\%
\begin{array}
[c]{c}%
u_{L}d_{L}(-g_{3L})\\
d_{L}d_{L}(-\sqrt{2}g_{3L})
\end{array}
\end{array}
\right)  $\\\hline
Vector Diquarks &  &  &  &  & \\
$DQ_{2\mu}$ & 3$^{\star}$ & 2 & -1/3 & $\left(
\begin{array}
[c]{c}%
1/3\\
-2/3
\end{array}
\right)  $ & $\left(
\begin{array}
[c]{c}%
d_{R}u_{L}(g_{2})\\
d_{R}d_{L}(-g_{2})
\end{array}
\right)  $\\
$\widetilde{DQ}_{2\mu}$ & 3$^{\star}$ & 2 & 5/3 & $\left(
\begin{array}
[c]{c}%
4/3\\
1/3
\end{array}
\right)  $ & $\left(
\begin{array}
[c]{c}%
u_{R}u_{L}(\tilde{g}_{2})\\
u_{R}d_{L}(-\tilde{g}_{2})
\end{array}
\right)  $\\\hline
\end{tabular}
\end{center}
\end{table}

For the present analysis, we consider the color $\mathbf{3}^{\ast}$ scalar
$DQ_{1}$ or $DQ_{3}^{0}$ diquarks coupled to $ud$ pairs, $\widetilde{DQ}_{1}$
or $DQ_{3}^{-}$ diquarks coupled to $dd$ pair and $\widetilde{DQ}_{1}^{\prime
}$ or $DQ_{3}^{+}$ diquarks coupled to $uu$\textit{\ }pair. The vector
diquarks $DQ_{2}^{1}$ and $\widetilde{DQ}_{2}^{2}$ of type $ud$, $DQ_{2}^{2}$
of type $dd$ and $\widetilde{DQ}_{2}^{1}$ of type $uu$ are considered. The
interaction between the diquark and quark pair is described by the effective
lagrangian (5) with different couplings.

\section{Decay Widths}

For the decay width calculation, we take the coupling as $g_{DQ}^{2}=g_{L}%
^{2}+g_{R}^{2}$ for each diquark type. For numerical results, we will use the
definition $g_{DQ}^{2}=4\pi\alpha_{DQ}$ when only one type of coupling assumed
to be nonzero. Diquarks decay into quark pairs, and the decay width
$\Gamma_{DQ}$ derived from the same lagrangian is
\begin{align}
\Gamma_{DQ}^{S}  &  ={\frac{F_{S}g_{DQ}^{2}m_{DQ}}{16\pi}}\simeq
25\,\text{GeV}({\frac{F_{S}m_{DQ}}{{1\,\text{TeV}}}})\;\;\;\text{\mbox{for}}%
\;\;\alpha_{DQ}=0.1\label{6}\\
\Gamma_{DQ}^{V}  &  ={\frac{F_{S}g_{DQ}^{2}m_{DQ}}{24\pi}}\simeq
17\text{GeV}({\frac{F_{S}m_{DQ}}{{1\,\text{TeV}}}})\;\;\;\text{\mbox{for}}%
\;\;\alpha_{DQ}=0.1 \label{7}%
\end{align}
where $F_{S}$ contains the color factor for a representation including the
statistical factors associated with the presence of identical fermions in the
final state, and $m_{DQ}$ is the mass of scalar ($S$) or vector ($V$) diquark.

\section{Signal and Background}

The signal for diquark production would clearly manifest itself in two jets
cross sections. The differential cross section for any type of scalar and
vector diquark resonant production can be written as follows%

\begin{align}
{\frac{d\hat{\sigma}^{S}}{d\hat{t}}}(q_{i}q_{j}  &  \rightarrow
DQ\rightarrow
q_{i}q_{j})={\frac{F{_{S}g}_{DQ}^{4}}{64\pi\lbrack(\hat{s}-m_{DQ}^{2}
)^{2}+(m_{DQ}\Gamma_{DQ}^{S})^{2}]}}\\
{\frac{d\hat{\sigma}^{V}}{d\hat{t}}}(q_{i}q_{j}  &  \rightarrow
DQ\rightarrow
q_{i}q_{j})={\frac{F{_{S}g}_{DQ}^{4}(\hat{t}^2+\hat{u}^2)}{16\pi\hat{s}^2\lbrack(\hat{s}-m_{DQ}^{2}
)^{2}+(m_{DQ}\Gamma_{DQ}^{V})^{2}]}.}
\end{align}
In the narrow width approximation ($\Gamma_{DQ}/m_{DQ}<0.1$), the cross
section of the $s-$channel diquark production can be obtained as%

\begin{align}
{\hat{\sigma}}_{R}^{S}(\widehat{s})  &
\simeq\frac{F{_{S}g}_{DQ}^{4}
\widehat{s}}{64m_{DQ}\Gamma_{DQ}^{S}}\delta(\widehat{s}-m_{DQ}^{2})\\
{\hat{\sigma}}_{R}^{V}(\widehat{s})  & \simeq\frac{F{_{S}g}_{DQ}^{4}
\widehat{s}}{48m_{DQ}\Gamma_{DQ}^{V}}\delta(\widehat{s}-m_{DQ}^{2})
\end{align}
where $\widehat{s}$ is the Mandelstam variable corresponding to the square of
center-of-mass energy for the subprocess.

\subsection{pp Collider}

\begin{figure}[ptb]
\begin{center}
\includegraphics[height=6cm,
width=8cm] {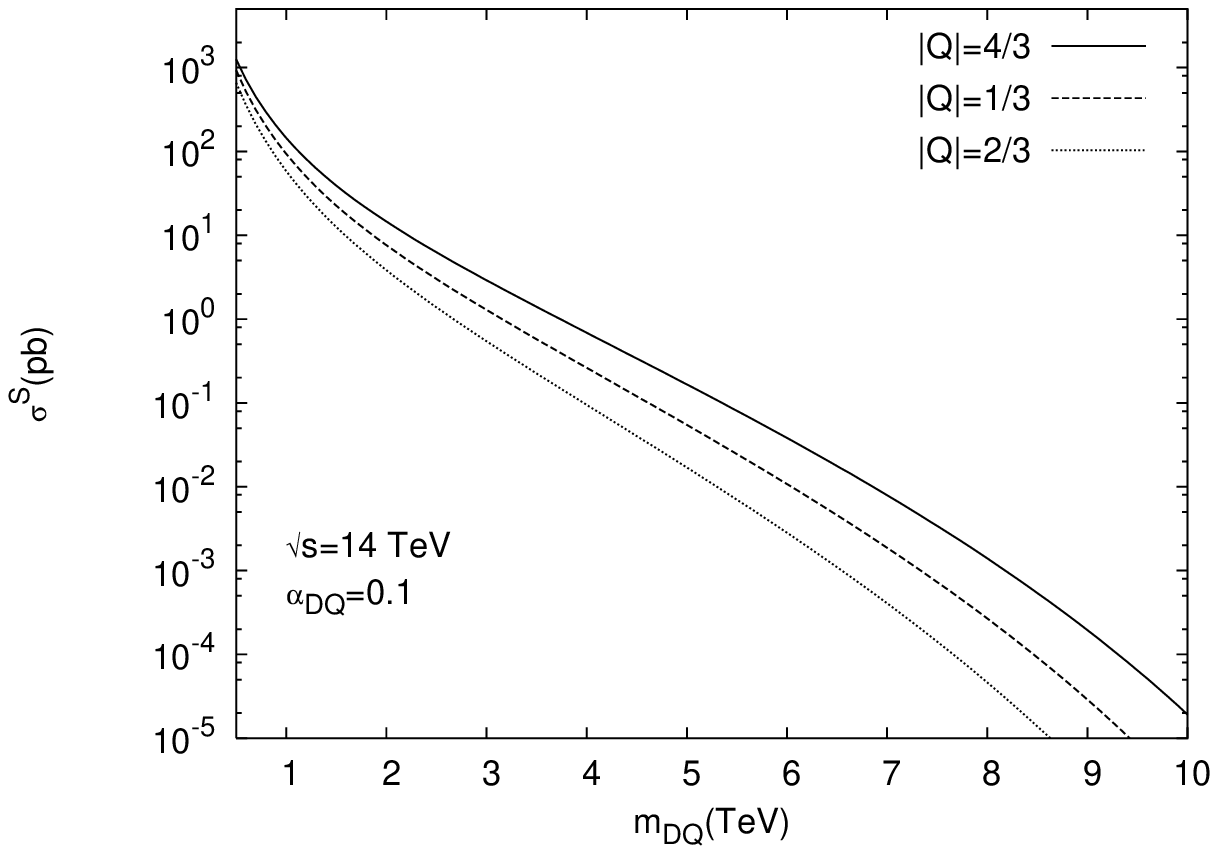}\includegraphics[height=6cm,
width=8cm] {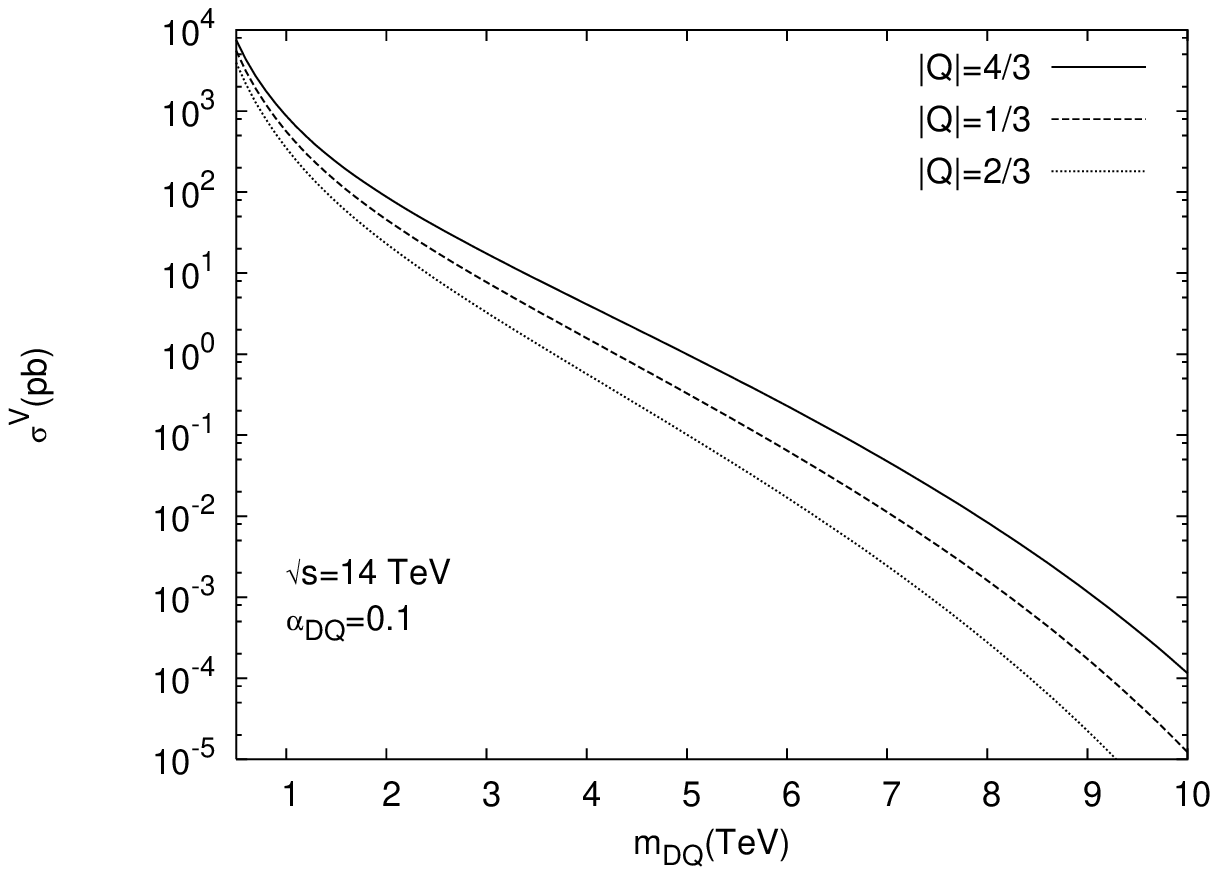}\\[0pt]\centerline{(a)\hspace{7cm}(b)}
\end{center}
\caption{Total cross sections in $pp$ collisions for a) scalar and b) vector
diquarks for different charges, with coupling strength $\alpha_{DQ}=0.1$,
depending on their masses. }%
\label{fig2}%
\end{figure}

The total cross section for the resonance production of scalar diquarks at
$pp$ collider is given by%

\begin{equation}
\sigma=\int_{m_{DQ}^{2}/s}^{1}\frac{dx}{x}f_{q/p}(x,Q_{p}^{2})f_{q^{\prime}%
/p}(m_{DQ}^{2}/xs,Q_{p}^{2})\widehat{{\sigma}}(\widehat{s}) \label{12}%
\end{equation}
where $f_{q/p}(x,Q_{p}^{2})$ and $f_{q^{\prime}/p}(x,Q_{p}^{2})$ correspond to
the quark distribution functions from the proton and we have used CTEQ5L
\cite{CTEQ5L} parametrization with $Q_{p}^{2}=\widehat{s}$. As a consequence
of this energy scan there will be quarks whose energies adequate for the
resonance production of diquarks. The cross section is plotted against the
diquark mass in Fig. \ref{fig2} (a-b) for LHC energy ($\sqrt{s}=14$ TeV) and
coupling $\alpha_{DQ}=0.1$. From these figures we find that diquarks with
charge $|Q|=4/3$ have the largest cross sections when compared to the other types.

The scalar and vector diquarks will decay via $DQ\rightarrow q_{i}q_{j}.$
Therefore, the relevant signal will be a pair of hard jets in the final state.
At the LHC energy, major QCD background processes contributing to two-jets
($2j$) final states and their integrated cross sections are given in Table
\ref{table3}.

\begin{figure}[ptb]
\begin{center}
\includegraphics[height=6cm,
width=8cm] {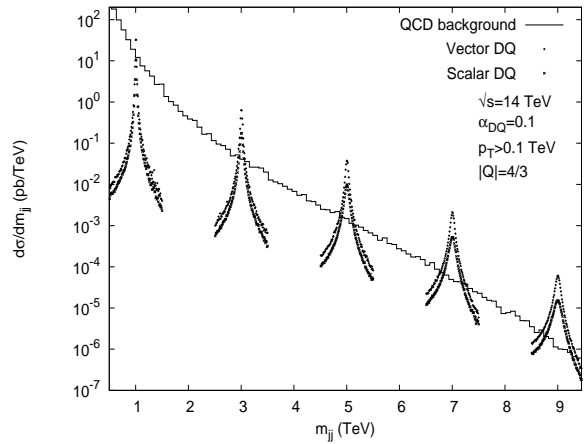}
\end{center}
\caption{Dijet invariant mass distributions for $pp\rightarrow2jX.$ Resonance
peaks are shown for scalar and vector diquark masses 1, 3, 5, 7, and 9 TeV for
comparison with smooth QCD backgrounds.}%
\label{fig3}%
\end{figure}

\begin{table}[ptb]
\caption{ The cross sections (in pb) for QCD backgrounds contributing to two
jets final states at parton level generated by CompHEP with various $p_{T}$
cuts.}%
\label{table3}
\begin{tabular}
[c]{lllllllll}\hline
Process &  & $p_{T}>0.1$ TeV &  & $p_{T}>0.5$ TeV &  & $p_{T}>1$ TeV &  &
$p_{T}>2$ TeV\\\hline
$gg\rightarrow gg$ &  & 6.3$\times10^{5}$ &  & 2.0$\times10^{2}$ &  &
2.3$\times10^{0}$ &  & 5.7$\times10^{-3}$\\
$qg\longrightarrow qg$ &  & 6.4$\times10^{5}$ &  & 4.8$\times10^{2}$ &  &
1.0$\times10^{1}$ &  & 5.7$\times10^{-2}$\\
$qq^{\prime}\rightarrow qq^{\prime}$ &  & 1.0$\times10^{5}$ &  &
1.8$\times10^{2}$ &  & 6.7$\times10^{0}$ &  & 8.8$\times10^{-2}$\\
$gg\rightarrow q\overline{q}$ &  & 2.4$\times10^{4}$ &  & 9.8$\times10^{0}$ &
& 1.0$\times10^{-1}$ &  & 2.9$\times10^{-4}$\\
$q\overline{q}\longrightarrow q^{\prime}\overline{q^{\prime}}$ &  &
1.6$\times10^{3}$ &  & 2.8$\times10^{0}$ &  & 1.3$\times10^{-1}$ &  &
1.1$\times10^{-3}$\\
$q\overline{q}\longrightarrow gg$ &  & 1.5$\times10^{3}$ &  & 2.5$\times
10^{0}$ &  & 6.7$\times10^{-2}$ &  & 8.5$\times10^{-4}$\\
Total &  & 1.4$\times10^{6}$ &  & 8.8$\times10^{2}$ &  & 1.9$\times10^{1}$ &
& 1.5$\times10^{-1}$\\\hline
\end{tabular}
\end{table}The values in Table \ref{table3} have been generated by CompHEP
program \cite{comphep} at parton level with various $p_{T}$ cuts on the jets.
It is clear that higher $p_{T}$ cuts reduce the background cross sections
significantly. These $p_{T}$ cuts can be translated into the rapidity cuts
according to the relation between the $p_{T}$ of a jet and the rapidity with
$p_{T}=m_{jj}/2\cosh y.$ Standard kinematic relations for the invariant mass
$m_{jj}$ and $p_{T}$ distributions of two-jet final states can be found in
\cite{Atag98,Arik02}. Figure \ref{fig3} shows the dijet invariant mass
distribution for the process $pp\rightarrow2j+X$ including the signal and the
QCD backgrounds at the LHC. For comparison signal peaks for scalar and vector
diquark masses $m_{DQ}=1,3,5,7,9$ TeV and $\alpha_{DQ}=0.1$ are shown on the
smooth background distribution.

\begin{figure}[ptb]
\begin{center}
\includegraphics[height=6cm,
width=8cm] {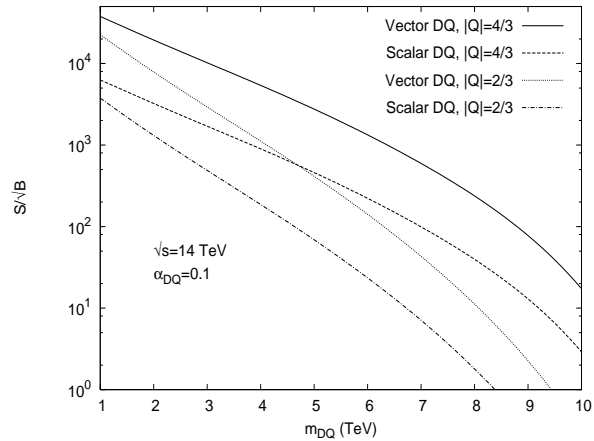}
\end{center}
\caption{The signal significances $S/\sqrt{B}$ for diquarks as a function of
diquark mass $m_{DQ}$ at the LHC. }%
\label{fig4}%
\end{figure}

In order to obtain the observability of diquarks at LHC we have calculated
signal (S) and background (B) event estimations for an integrated luminosity
of $10^{5}$ pb$^{-1}$. The signal generated by a diquark of mass $m_{DQ}$ and
decay rate $\Gamma_{DQ}$ is calculated integrating the differential cross
section in the two-jet invariant mass interval $m_{DQ}-\Gamma_{DQ}%
<m_{jj}<m_{DQ}+\Gamma_{DQ}$ which embraces approximately $95\%$ of the events
around the resonance. For a realistic analysis of the background events we
take into account the finite energy resolution of the LHC-ATLAS hadronic
calorimeter \cite{ATLAS} as $\delta E/E=0.5/\sqrt{E}+0.03$ for jets with
$|y|<3.$ The corresponding two-jet invariant mass resolution is given
approximately by $\delta m_{jj}=0.5\sqrt{m_{jj}}+0.02m_{jj}$. The background
is calculated by integrating the cross sections in the range $m_{DQ}-\Delta
m<m_{jj}<m_{DQ}+\Delta m$ with $\Delta m=\max(\Gamma_{DQ},\delta m_{jj})$. The
significance of signal over background is defined as $S/\sqrt{B}.$ In Fig.
\ref{fig4} we present $S/\sqrt{B}$ as a function of the diquark mass for the
scalar and vector diquarks with charges $|Q|=4/3$ and $2/3.$

If we take at least $25$ signal events and $S/\sqrt{B}\geq5$ as
discovery criteria, scalar (vector) diquarks with charge $|Q|=2/3$ can be
observed up to $7.5$ ($8.5$) TeV. For the diquarks with charge $|Q|=4/3$ it is
possible to cover mass ranges up to $9.5$ ($10)$ TeV at the LHC with
$L_{int}=10^{5}$ pb$^{-1}$. For this luminosity, $10^{7}$ scalar diquark
events/year and $10^{8}$ vector diquark events/year are expected for
$m_{DQ}=1$ TeV. Our results show that even for much lower coupling constants
as $10^{-3}$, diquarks should be seen at the LHC.

\subsection{ep collider}

In order to calculate the total cross section for diquark production due to
resolved photon process at $ep$ colliders, we use the formula%

\begin{equation}
\sigma=\int_{m_{DQ}^{2}/s}^{1}\frac{dx}{x}\int_{x}^{1}\frac{dy}{y}f_{\gamma
/e}(y)f_{q^{\prime}/\gamma}(x/y,Q_{\gamma}^{2})f_{q/p}(m_{DQ}^{2}/xs,Q_{p}%
^{2})\widehat{{\sigma}}(\widehat{s}) \label{13}%
\end{equation}
where $f_{\gamma/e}(y)$ \cite{Florian99} and $f_{q^{\prime}/\gamma
}(x/y,Q_{\gamma}^{2})$ \cite{Gluck92} corresponds to the photon energy
spectrum and the probability density of finding a quark ($q\prime$) carrying a
fraction $x/y$ of the energy of the electrons in the initial beams,
respectively. In (13) the integrals over the momentum fractions $x$ and $y$
are the mathematical representation of an energy scan performed by the partons
coming from photons and protons. We use the photon spectrum $f_{\gamma/e}(y)$
resulting from bremsstrahlung \cite{Florian99} with the electron beam energy
$E_{e}$%

\begin{equation}
f_{\gamma/e}(y)=\frac{\alpha}{2\pi}\left[  \frac{1+(1-y)^{2}}{y}\ln
\frac{4E_{e}^{2}(1-y)^{2}}{m_{e}^{2}y^{2}}+2m_{e}^{2}y\left(  \frac{1}%
{4E_{e}^{2}(1-y)}-\frac{1-y}{m_{e}^{2}y^{2}}\right)  \right]  .
\end{equation}

\begin{figure}[ptb]
\begin{center}
\includegraphics[height=6cm,
width=8cm] {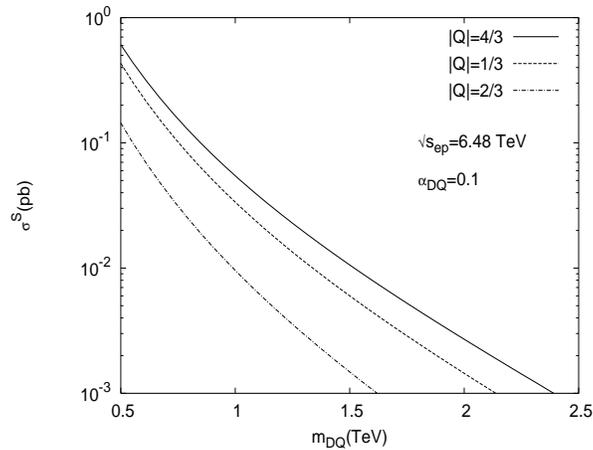}
\end{center}
\caption{The total cross sections for the process $ep\rightarrow
DQ+X\rightarrow2j+X$ at $\sqrt{s_{ep}}=6.48$ TeV depending on the mass of
scalar diquarks.}%
\label{fig5}%
\end{figure}

We present the cross sections for scalar diquarks against their masses for
coupling $\alpha_{DQ}=0.1$ and charges $|Q|=4/3,1/3$ and $2/3$ using the
bremsstrahlung process in Fig. \ref{fig5} at CLIC$\otimes$LHC based $ep$
colliders with $\sqrt{s_{ep}}=6.48$ TeV.

\begin{figure}[ptb]
\begin{center}
\includegraphics[height=6cm,
width=8cm] {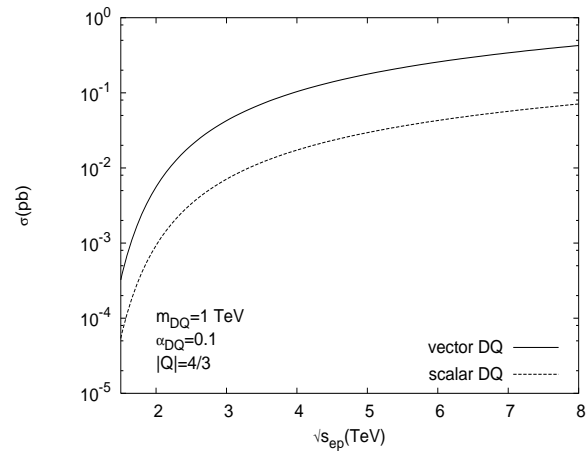}
\end{center}
\caption{The production cross sections $\sigma(ep\rightarrow DQ+X\rightarrow
2j+X)$ for scalar and vector diquarks with charge $|Q|=4/3$ and coupling
$\alpha_{DQ}=0.1$ depending the center of mass energy of the $ep$ colliders.}%
\label{fig6}%
\end{figure}

In order to see the potentials of possible options of $ep$
colliders with different center of mass energies, we plot diquark production
cross sections depending on the $\sqrt{s_{ep}}$. As seen from the Fig.
\ref{fig6} vector diquarks have larger cross sections than the scalar for the
considered center of mass energy region. In principle, the vector and scalar
type diquarks can be easily distinguished by the angular distribution of
produced jets, the electric charge of diquarks can be determined by the
interactions with photon.

In the resonant production mechanism of diquarks with resolved
photon process in collision $ep\rightarrow DQ+X\rightarrow2j+X$, most of the
partcles in the final state are expected to be lost in the beam pipe, the two
exceptions being the quarks resulting from diquark decay. As a result, the
relevant signal is a $p_{T}$ balanced coplanar pair of jets and it will
contain two hard jets in the final state. All the relevant interactions with
the diquarks are implemented into the CompHEP \cite{comphep} program. The
decay width of diquarks and the cross sections for signal and background
processes are calculated with this program.

In order to study contributing backgrounds we need to take into account all
the diagrams contributing to two-jets final states. The process $ep\rightarrow
2j+X$ has potentially large QCD background. At the $ep$ colliders, major
processes contributing to two-jets ($2j$) final states and their integrated
cross sections are given in Table \ref{table4}.

\begin{figure}[ptb]
\begin{center}
\includegraphics[height=6cm,
width=8cm] {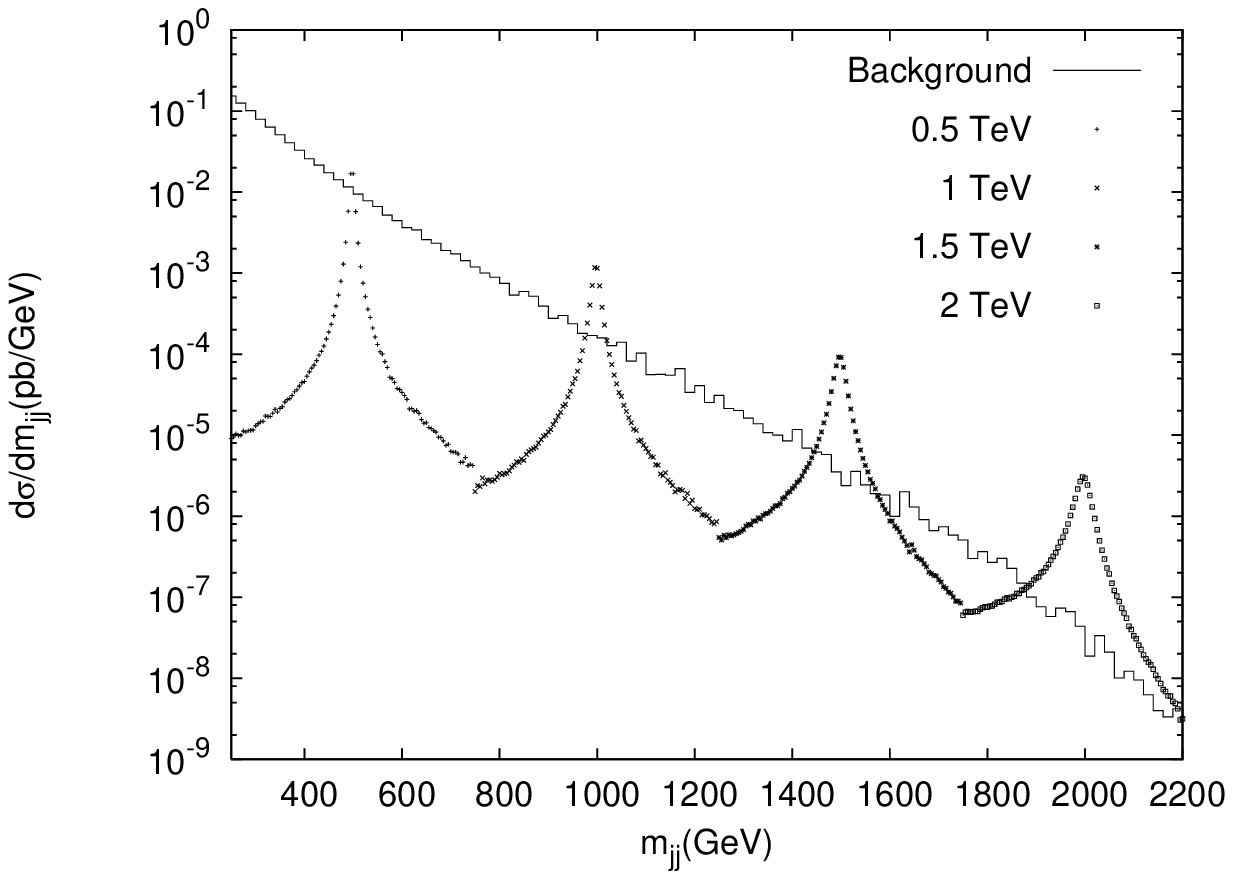}
\end{center}
\caption{The invariant mass distributions for the QCD background with the
inclusion of scalar diquark signals at $ep$ collisons. }%
\label{fig7}%
\end{figure}

\begin{table}[ptb]
\caption{ The major processes contributing to two jet ($jj$) final states with
a cut $p_{T}>100$ GeV at the $ep$ collisions with different center of mass
energy $\sqrt{s_{ep}}$.}%
\label{table4}
\begin{center}%
\begin{tabular}
[c]{lllllll}\hline
Collider &  & ILC$\otimes$LHC &  & \multicolumn{3}{l}{CLIC$\otimes$LHC}\\
$\sqrt{s_{ep}}$(TeV) &  & 2.64 &  & 3.74 &  & 6.48\\\hline
$\gamma\overline{q}\rightarrow g\overline{q}$ &  & $2.631\times10^{0}$ &  &
$5.064\times10^{0}$ &  & $1.234\times10^{1}$\\
$\gamma q\rightarrow gq$ &  & $8.275\times10^{0}$ &  & $1.257\times10^{1}$ &
& $2.290\times10^{1}$\\
$\gamma g\rightarrow q\overline{q}$ &  & $1.522\times10^{1}$ &  &
$3.248\times10^{1}$ &  & $9.233\times10^{1}$\\
Total &  & $2.613\times10^{1}$ &  & $5.011\times10^{1}$ &  & $1.276\times
10^{2}$\\\hline
\end{tabular}
\end{center}
\end{table}The values in Table \ref{table4} have been generated by CompHEP at
parton level for transverse momentum cut of the jets $p_{T}>100$ GeV.
Fig.~\ref{fig7} shows jet-jet invariant mass distribution for the process
$ep\rightarrow2jX$ at CLIC$\otimes$LHC based $ep$ collider with $\sqrt{s_{ep}%
}=6.48$ TeV.

\begin{figure}[ptb]
\begin{center}
\includegraphics[height=6cm,
width=8cm] {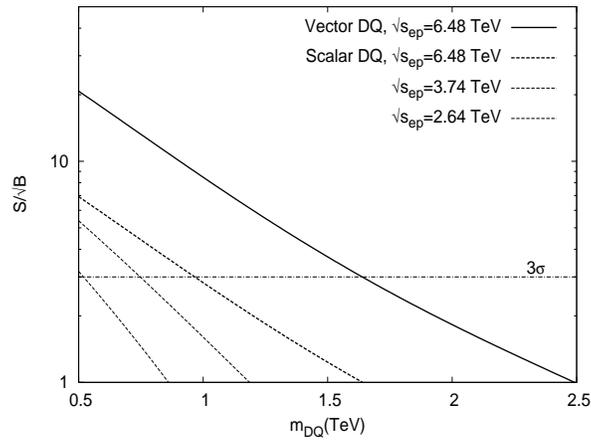}
\end{center}
\caption{The signal significances $S/\sqrt{B}$ as a function of scalar and
vector diquark mass $m_{DQ}$ at different $ep$ collider energies. }%
\label{fig8}%
\end{figure}

In our analysis, we determine the expected significance of the signal over
background for collider parameters given in Table \ref{table1} for $ep$
colliders. It is convenient to collect the data in two-jet invariant mass bins
since the signal is concentrated in a small region of the invariant mass
spectrum. Therefore, we introduced the cut $\mid m_{jj}-$ $m_{DQ}\mid\leq25$
GeV which is efficient only for diquark masses between 0.5 and 1 TeV. For
heavier diquarks rather broad resonances are expected and the cut $\mid
m_{jj}-$ $m_{DQ}\mid\leq50$ GeV becomes suitable for $1$ TeV$<m_{DQ}<2.5$ TeV.
This range of the invariant mass will embrace approximately $95\%$ of the
signal events around the diquark resonance. For a realistic analysis we have
estimated background events for $ep$ colliders with the integrated luminosity
of $10^{3}$ pb$^{-1}$ taking into account the energy resolution of the
hadronic calorimeter. The statistical significances $S/\sqrt{B}$ of diquark
signal over background are shown in Fig. \ref{fig8} for ILC$\otimes$LHC and
CLIC$\otimes$LHC based $ep$ colliders. From Fig. \ref{fig8}, we find that
scalar (vector) diquarks can be seen up to $1$ ($1.6$) TeV at the $ep$
collider with $\sqrt{s}=6.48$ TeV. A slightly lower limits can be obtained for
the other $ep$ options.

If we take into account the integrated luminosity $L=10^{3}$ pb$^{-1}$ for the
CLIC$\otimes$LHC based $ep$ collider with $\sqrt{s}=6.48$ TeV, $44$ scalar
diquark events and $264$ vector diquark events/year are expected for
$m_{DQ}=1$ TeV and $\alpha_{DQ}=0.1$.

\subsection{e$^{+}$e$^{-}$ collider}

The cross section for the production of a pair of jets through the two
resolved photon process is given by%

\begin{equation}
\sigma=\int_{x_{\min}}^{x_{\max}}dx\int_{y_{\min}}^{y_{\max}}dyf_{q/e}%
(x,Q_{\gamma}^{2})f_{q^{\prime}/e}(y,Q_{\gamma}^{2})\widehat{{\sigma}%
}(\widehat{s}) \label{17}%
\end{equation}
here $f_{q/e}(x,Q_{\gamma}^{2})$ and $f_{q^{\prime}/e}(y,Q_{\gamma}^{2})$
correspond to the probability density of finding a quark $q$ and $q^{\prime}$
carrying a fraction $x$ and $y$ of the energy of the electrons and positrons
in the initial beams, respectively. In (15) the integrals over the momentum
fractions $x$ and $y$ are the mathematical representation of an energy scan
performed by the partons coming from electrons and positrons. The lower limits
of the integrations guarantee the energy required for the production of
quarks. As a consequence of this energy scan there will be quarks whose
energies adequate for the resonance production of diquarks. For the
probability $f_{q/e}(x,Q_{\gamma}^{2})$ resulting from the convolution of the
spectrum of photons $f_{\gamma/e}(x_{2})$ converted from the initial beams and
photonic parton distribution function $f_{q/\gamma}(x_{1},Q_{\gamma}^{2})$
\cite{Gluck92}, we use%

\begin{align}
f_{q/e}(x,Q_{\gamma}^{2})  &  =\int_{x_{2\min}}^{x_{2\max}}dx_{2}%
\int_{x_{1\min}}^{x_{1\max}}dx_{1}f_{\gamma/e}(x_{2})f_{q/\gamma}%
(x_{1},Q_{\gamma}^{2})\delta(x_{1}x_{2}-x)\\
&  =\int_{x}^{1}\frac{dx_{2}}{x_{2}}f_{\gamma/e}(x)f_{q/\gamma}(\frac{x}%
{x_{2}},Q_{\gamma}^{2})
\end{align}
For the relevant $Q_{\gamma}^{2}$ value entering the parton distribution in
the photon, we have taken $Q_{\gamma}=\sqrt{\widehat{s}}/2.$

\begin{figure}[ptb]
\begin{center}
\includegraphics[height=6cm,
width=8cm] {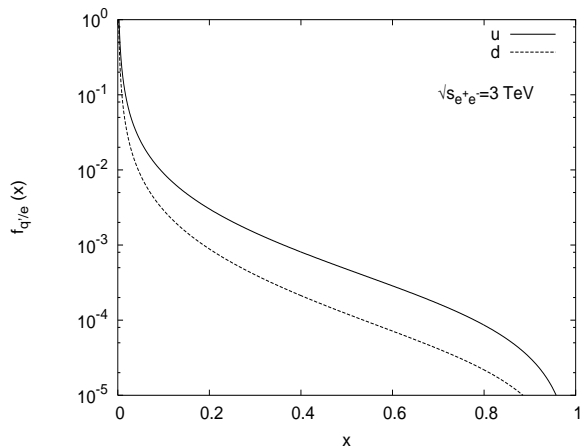}
\end{center}
\caption{The probability density for up ($u)$ and down ($d)$ quarks within
electron beam for bremsstrahlung photons.}%
\label{fig9}%
\end{figure}

In Fig. \ref{fig9}, the quark probability density function $f_{q/e}%
(x,Q_{\gamma}^{2})$ within electron beam are presented for bremsstrahlung
photons at beam energy $E_{e}=1.5$ TeV.

\begin{figure}[ptb]
\begin{center}
\includegraphics[height=6cm,
width=8cm] {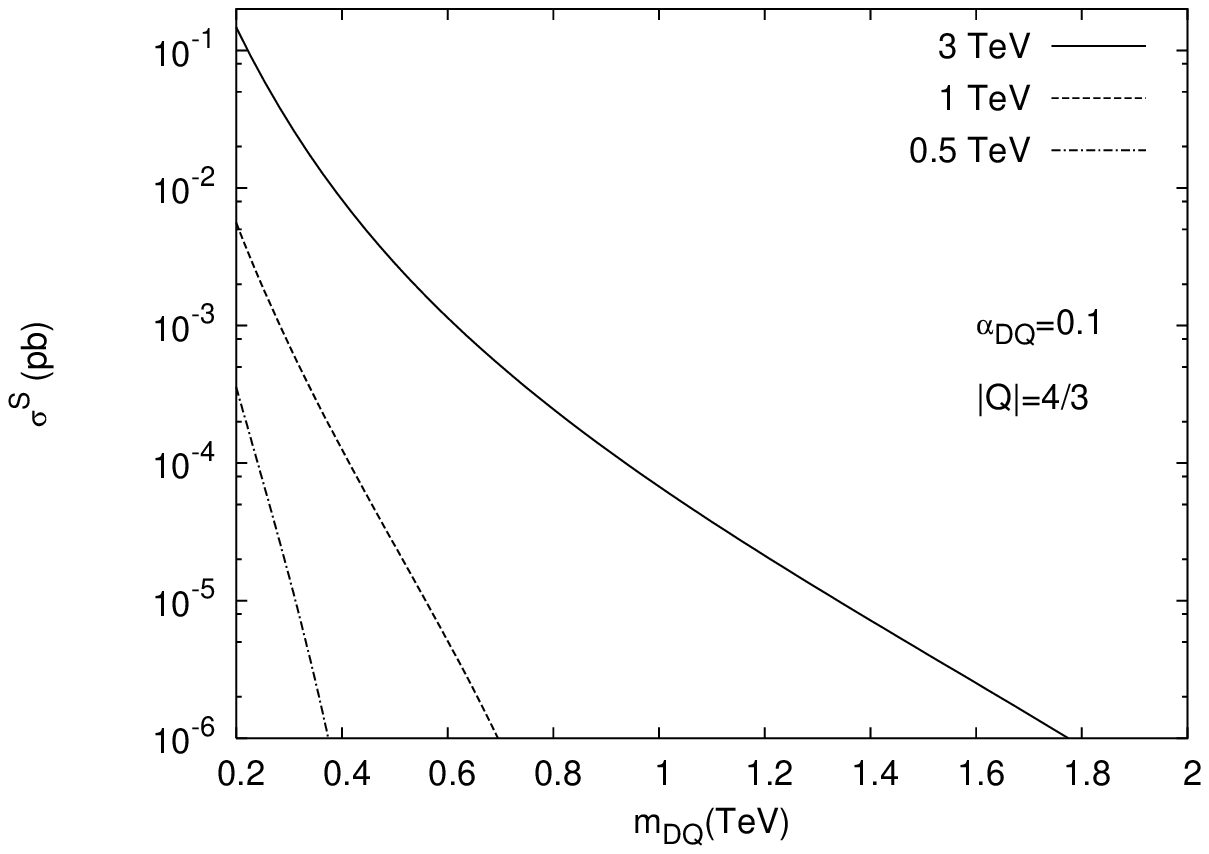}\includegraphics[height=6cm,
width=8cm] {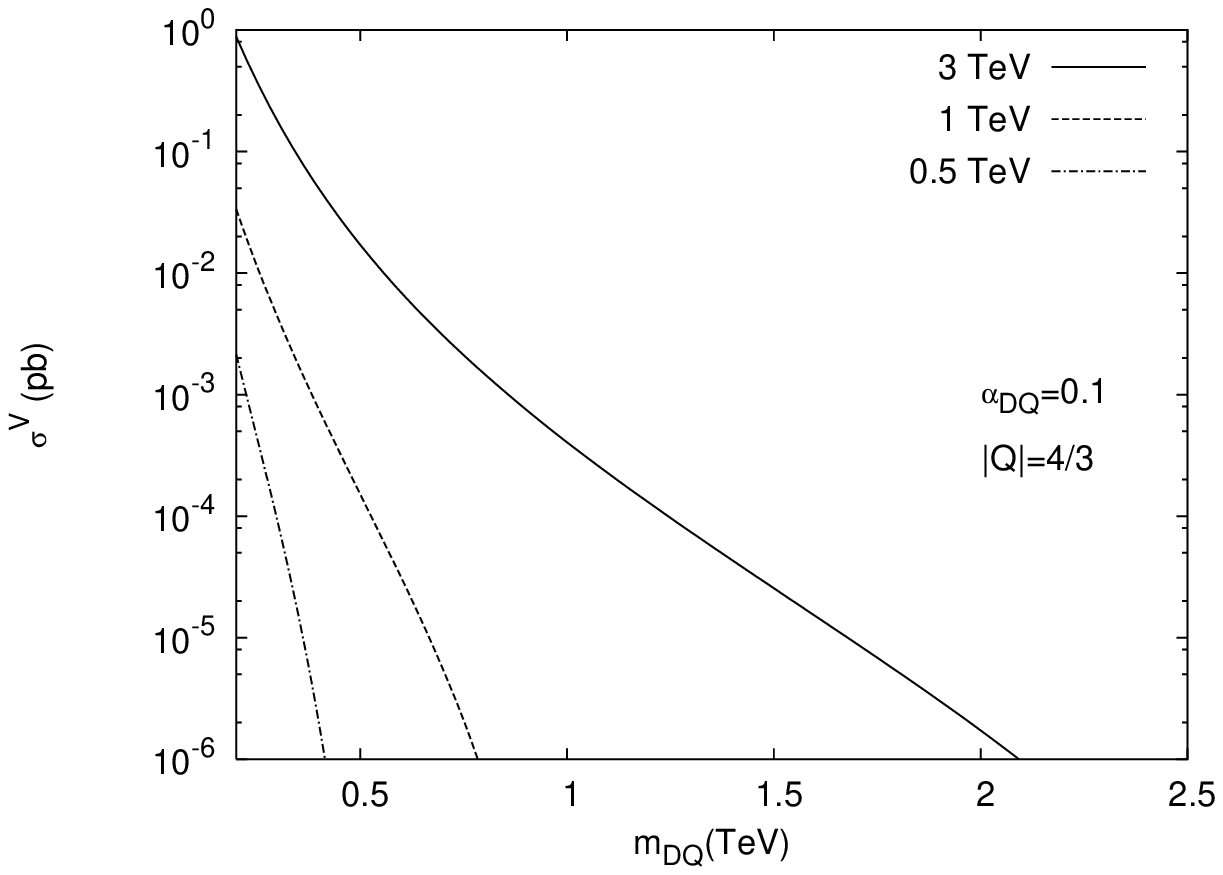}\newline\centerline{(a)\hspace{7cm}(b)}
\end{center}
\caption{The production cross sections for (a) scalar diquark and (b) vector
diquark with charge $|$$Q|=4/3$ at $e^{+}e^{-}$ colliders with $\sqrt
{s_{e^{+}e^{-}}}=0.5,1$ and $3$ TeV, where the coupling $\alpha_{DQ}=0.1$.}%
\label{fig10}%
\end{figure}

Here, we show only $u$ and $d$ quark distributions in photon since
we deal with only the first family diquarks. In Fig.~\ref{fig10} (a-b), the
production cross section versus scalar and vector diquark mass is plotted for
$e^{+}e^{-}$ with the center of mass energy $\sqrt{s_{e^{+}e^{-}}}=0.5,1$ and
$3$ TeV$.$

The two-jet background receives contributions from the annihilation process
$e^{+}e^{-}\rightarrow\gamma,Z\rightarrow q\overline{q}$ and the hard two
photon processes. The two photon processes can be of type direct, one resolved
and two resolved. The last two processes, despite being higher order in
$\alpha_{s},$ contribute significantly at lower energies. The spectrum of
initial state radiation (ISR, photon radiation from the incoming electrons and
positrons) contains the ISR scale inherent to the process under consideration
\cite{Kuraev85}. In our calculations, this spectrum is taken into account
using the CompHEP program with the convolution of beamstrahlung spectra.
Beamstrahlung is the process of energy loss by the incoming electron/positron
in the field of the positron/electron bunch moving in the opposite direction.
Beamstrahlung spectrum is an attribute of the linear collider design and
depends on the bunch geometry, bunch charge and the collision energy. The
photon spectrum resulting from the electromagnetic interaction between the
electron and positron beams in the intersection region is described by a more
complicated expression \cite{Chen92}. The beamstrahlung parameters, $\Upsilon$
and $N_{\gamma}$, which in their turn are determined by the bunch design of a
linear collider:
\begin{equation}
\Upsilon={\frac{5\alpha NE_{e}}{6m_{e}^{3}\sigma_{z}(\sigma_{x}+\sigma_{y})}%
}\quad,\qquad N_{\gamma}={\frac{25\alpha^{2}N}{12m_{e}(\sigma_{x}+\sigma_{y}%
)}}{\frac{1}{\sqrt{1+\Upsilon^{2/3}}}}%
\end{equation}
where $N$ is the number of particles in the bunch; $E_{e}$ and $m_{e}$ are the
energy and mass of the electron, respectively; $\sigma_{x},\sigma_{y}$ and
$\sigma_{z}$ are the average sizes of the particle bunches. We presented the
beamstrahlung parameters in Table ~\ref{table5} and their effects in the
background calculations in Table ~\ref{table6}.

\begin{table}[ptb]
\caption{Collider parameters relevant for the calculation of beamstrahlung.
$\Upsilon$ is the beamstrahlung parameter and $N_{\gamma}$ is the average
number of photons per electron. }%
\label{table5}
\begin{tabular}
[c]{lllllll}\hline
Collider &  & ILC &  & \multicolumn{3}{l}{\qquad\quad CLIC}\\
parameter &  & 500 GeV &  & 1 TeV &  & 3 TeV\\\hline
N(10$^{10}$) &  & 2 &  & 0.4 &  & 0.4\\
$\sigma_{x}($nm$)$ &  & 655 &  & 115 &  & 43\\
$\sigma_{y}($nm$)$ &  & 5.7 &  & 1.75 &  & 1\\
$\sigma_{z}(\mu$m$)$ &  & 300 &  & 30 &  & 30\\
$\Upsilon$ &  & 0.045 &  & 1.014 &  & 8.068\\
$N_{\gamma}$ &  & 1.22 &  & 1.04 &  & 1.74\\\hline
\end{tabular}
\end{table}

\begin{table}[ptb]
\caption{Background calculation with the ISR and beamstrahlung effects. The
numbers are the cross sections in pb for the process $e^{+}e^{-}%
\rightarrow\gamma,Z\rightarrow q\bar{q}$ with $p_{T}^{j}>20$ GeV ($p_{T}%
^{j}>100$ GeV). }%
\label{table6}
\begin{tabular}
[c]{lllllll}\hline
Collider &  & ILC &  & \multicolumn{3}{l}{\qquad\quad CLIC}\\
parameter &  & 500 GeV &  & 1 TeV &  & 3 TeV\\\hline
no ISR &  & 2.22(1.96) &  & 0.546(0.672) &  & 0.0604(0.0602)\\
with ISR &  & 8.16(2.11) &  & 1.96(0.629) &  & 0.276(0.0829)\\
with ISR+beamstrahlung &  & 8.38(2.16) &  & 4.30(1.04) &  &
67.55(1.39)\\\hline
\end{tabular}
\end{table}

\begin{figure}[ptb]
\begin{center}
\includegraphics[height=6cm,
width=8cm] {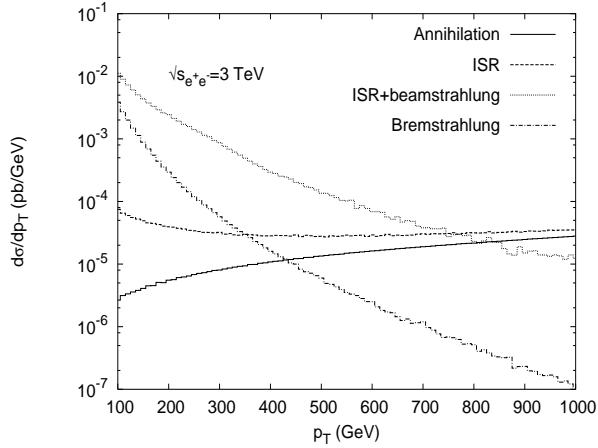}
\end{center}
\caption{The transverse momentum distributions of the jet.}%
\label{fig11}%
\end{figure}

We determine the expected significance of the signal over the background for
collider parameters as given in Table \ref{table1}, for ILC operating at 500
GeV, CLIC operating at 1 and 3 TeV. For the realistic analysis of the
background we consider the generic hadron calorimeters with the two-jet mass
resolution $\delta m_{jj}=0.5\sqrt{m_{jj}}+0.03m_{jj}.$ In order to reduce the
background we vetoed events exhibiting $Z$-bosons decaying into two-jets
through the cut $m_{jj}>100$ GeV. In Fig. \ref{fig11}, we presented the
contributions from both the annihilation process with/without
ISR+beamstrahlung and two-photon processes to dijet cross sections at
$\sqrt{s_{e^{+}e^{-}}}=3$ TeV.

\begin{figure}[ptb]
\begin{center}
\includegraphics[height=6cm,
width=8cm] {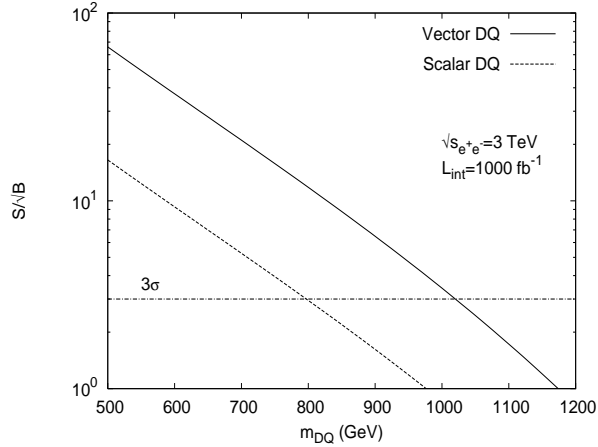}
\end{center}
\caption{The scalar and vector diqurk signal significance as a function of
their masses.}%
\label{fig12}%
\end{figure}

We conclude that signal significances scale with the collider luminosity as
$\sqrt{L},$ and higher luminosities are desirable for a good signal over
background discrimination. In determining the background events we considered
some cuts on transverse momentum $p_{T}>100$ GeV of the jets. In Fig.
\ref{fig12}, we present the signal significance that can be reached at
$e^{+}e^{-}$ collider with at $\sqrt{s_{e^{+}e^{-}}}=3$ TeV.

We take the indication of signal $S/\sqrt{B}\geq3$ and find that vector
diquarks can be probed up to $1$ TeV at $e^{+}e^{-}$ collider with
$\sqrt{s_{e^{+}e^{-}}}=3$ TeV. For the $e^{+}e^{-}$ colliders running at the
center of mass energy of $1$ TeV we can search for scalar diquarks up to $325$
GeV and vector diquarks up to $375$ GeV.

\section{Conclusion}

The resonant production of scalar and vector diquarks at LHC have large cross
section. With reasonable cuts, it may be possible to cover mass ranges up to
$10$ TeV for coupling $\alpha_{D}=0.1$. For smaller couplings as $\alpha
_{D}=10^{-3}$, it is still possible to probe diquarks up to the mass of $4$
TeV at an integrated luminosity $L=10^{2}$fb$^{-1}$. We find that vector
diquarks can be seen up to $1.6$ TeV at the $ep$ collider with $\sqrt{s}=6.48$
TeV and a slightly lower limits can be obtained for scalar diquarks. The two
hard photon processes prevail at energies below $\sqrt{s}/2$ while the
annihilation prevails at higher energies. Therefore, the attainable mass
limits for diquarks are smaller than $\sqrt{s}/2$ at linear colliders with the
parameters given in Table \ref{table1}.

A limited number of measurements of diquark properties can be carried out at
the hadron colliders. The future high energy linear $e^{+}e^{-}$ colliders and
linac-ring type $ep$ colliders based on the hadron and lepton colliders would
complement the measurements performed at hadron colliders. The resolved photon
contributions to the search for new physics becomes important for forthcoming
$ep$ and $e^{+}e^{-}$ collisions.

\begin{acknowledgments}
This work is partially supported by the Turkish State Planning Organization
(DPT) under the grant No 2002K120250 and 2003K120190.
\end{acknowledgments}

\end{document}